\documentclass[11pt]{article}

\usepackage{amsmath, amsthm}
\usepackage{amsmath, amsfonts}
\usepackage{amsmath, amssymb}
\usepackage{amsmath}
\usepackage{graphics}
\usepackage{graphicx}
\usepackage{color}
\usepackage{hyperref}
\usepackage[all]{xy}

\newtheorem{definition-lemma}[theorem]{Definition/Lemma}
\newtheorem{definition-explanation}[theorem]{Definition/Explanation}
\newtheorem{explanation-definition}[theorem]{Explanation/Definition}
\newtheorem{definition-fact}[theorem]{Definition/Fact}
\newtheorem{definition-notation}[theorem]{Definition/Notation}
\newtheorem{definition-conjecture}[theorem]{Definition/Conjecture}

\newtheorem{lemma-definition}[theorem]{Lemma/Definition}

\newtheorem{remark-notation}[theorem]{\it Remark/Notation}

\newtheorem{application-lemma}[theorem]{Application/Lemma}

\newtheorem{example-definition}[theorem]{Example/Definition}

\newtheorem{definition-prototype}[theorem]{Definition-Prototype}

\numberwithin{equation}{subsection}

\newtheorem{stheorem}{Theorem}[section]
\newtheorem{sdefinition}[stheorem]{Definition}
\newtheorem{sdefinition-lemma}[stheorem]{Definition/Lemma}
\newtheorem{sdefinition-explanation}[stheorem]{Definition/Explanation}
\newtheorem{sexplanation-definition}[stheorem]{Explanation/Definition}
\newtheorem{sdefinition-fact}[stheorem]{Definition/Fact}
\newtheorem{sdefinition-notation}[stheorem]{Definition/Notation}
\newtheorem{sdefinition-conjecture}[stheorem]{Definition/Conjecture}
\newtheorem{slemma}[stheorem]{Lemma}
\newtheorem{slemma-definition}[stheorem]{Lemma/Definition}
\newtheorem{sproposition}[stheorem]{Proposition}
\newtheorem{scorollary}[stheorem]{Corollary}
\newtheorem{sremark}[stheorem]{\it Remark}
\newtheorem{sremark-notation}[stheorem]{\it Remark/Notation}

\newtheorem{sapplication-lemma}[stheorem]{Application/Lemma}

\newtheorem{sexample-definition}[stheorem]{Example/Definition}

\newtheorem{snotation}[stheorem]{Notation}

\newtheorem{sdefinition-prototype}[stheorem]{Definition-Prototype}

\newtheorem{ssdefinition-lemma}[sstheorem]{Definition/Lemma}
\newtheorem{ssdefinition-explanation}[sstheorem]{Definition/Explanation}
\newtheorem{ssexplanation-definition}[sstheorem]{Explanation/Definition}
\newtheorem{ssdefinition-fact}[sstheorem]{Definition/Fact}
\newtheorem{ssdefinition-notation}[sstheorem]{Definition/Notation}
\newtheorem{ssdefinition-conjecture}[sstheorem]{Definition/Conjecture}

\newtheorem{sslemma-definition}[sstheorem]{Lemma/Definition}

\newtheorem{ssremark-notation}[sstheorem]{\it Remark/Notation}

\newtheorem{ssapplication-lemma}[sstheorem]{Application/Lemma}

\newtheorem{ssexample-definition}[sstheorem]{Example/Definition}

\newtheorem{ssdefinition-prototype}[sstheorem]{Definition-Prototype}

\newcommand{\Comm}{\mbox{\it Comm}\,}

\newcommand{\CSWZ}{\mbox{\scriptsize\it CS/WZ}\,}
 
\newcommand{\DBI}{\mbox{\it\scriptsize DBI}\,}

\newcommand{\End}{\mbox{\it End}\,}

\newcommand{\Endsheaf}{\mbox{\it ${\cal E}\!$nd}\,}

\newcommand{\GL}{\mbox{\it GL}}

\newcommand{\Real}{\mbox{\it Re}\,}

\newcommand{\Sym}{\mbox{\it Sym}}

\newcommand{\SymDet}{\mbox{\it SymDet}\,}

\newcommand{\Tr}{\mbox{\it Tr}\,}

\newcommand{\pr}{\mbox{\it pr}}

\newcommand{\redscriptsize}{\mbox{\scriptsize\rm red}\,}
\newcommand{\redtiny}{\mbox{\tiny\rm red}\,}

\newcommand{\odotwedge}{\stackrel{\mbox{\tiny $\odot$}}{\wedge}}

\newcommand{\LARGEdot}{\mbox{\LARGE $\cdot$}}

\newcommand{\tinybullet}{\raisebox{.2ex}{\tiny $\bullet$}}							
							
\begin{document}

\enlargethispage{24cm}

\begin{titlepage}

$ $

\vspace{-1.5cm} 

\noindent\hspace{-1cm}
\parbox{6cm}{\small November 2016}\
   \hspace{7cm}\
   \parbox[t]{6cm}{yymm.nnnnn [hep-th] \\
                D(13.2.1): massless condition 
				}

\vspace{2cm}

\centerline{\large\bf
 More on the admissible condition on differentiable maps
  $\varphi: (X^{\!A\!z},E;\nabla)\rightarrow Y$}
\vspace{1ex}
\centerline{\large\bf
 in  the construction of the non-Abelian Dirac-Born-Infeld action $S_{DBI}(\varphi,\nabla)$}

\bigskip

\vspace{3em}

\centerline{\large
  Chien-Hao Liu
            \hspace{1ex} and \hspace{1ex}
  Shing-Tung Yau
}

\vspace{6em}

\begin{quotation}
\centerline{\bf Abstract}

\vspace{0.3cm}

\baselineskip 12pt  
{\small
 In D(13.1) (arXiv:1606.08529 [hep-th]),
  we introduced an admissible condition on differentiable maps
  $\varphi: (X^{\!A\!z}, E;\nabla)\rightarrow Y$
  from an Azumaya/matrix manifold $X^{\!A\!z}$ (with the fundamental module $E$)
   with a connection $\nabla$ on $E$ to a manifold $Y$
  in order to resolve a pull-push issue in the construction of a non-Abelian-Dirac-Infeld action $S_{DBI}$
  for $(\varphi,\nabla)$ and to render $\nabla$ massless from the aspect of open strings.
 The admissible condition ibidem consists of two parts: Condition (1) and Condition (2).
 In this brief note$^{\ast}$, we examine these two conditions in more detail
  and bring their geometric implications on
    $(\varphi,\nabla)$ and the full action $S_{DBI}(\varphi,\nabla)+S_{CS/WZ}(\varphi,\nabla)$
	more transparent.
 In particular, we show that Condition (1)  alone already implies masslessness of $\nabla$ from open-string aspect;
  and that the additional Condition (2) implies a decoupling of the nilpotent fuzzy cloud of $\varphi(X^{\!A\!z})$
  to the dynamics of $(\varphi, \nabla)$.
 We conclude with a refined definition of admissible $(\varphi,\nabla)$ and
   a remark on the anomaly factor in the integrand of the Chern-Simons/Wess-Zumino term
   $S_{CS/WZ}(\varphi,\nabla)$ for D-branes based on the current study.
 } 
\end{quotation}

\vspace{8em}

\baselineskip 12pt
{\footnotesize
} 

 \bigskip

\noindent {\small MSC number 2010: 81T30, 53C05; 16S50, 14A22, 35R01.
} 

\bigskip

\baselineskip 10pt
{\scriptsize
\noindent{\bf Acknowledgements.}
We thank
 Andrew Strominger, Cumrun Vafa
   for influence to our understanding of strings, branes, and gravity.
C.-H.L.\ thanks in addition
 Brendan McLellan, Peter Smilie, Chenglong Yu, Boyu Zhang
   for discussions that lead to Lemma 2.1;
 Joachim Jelisiejew
   for a highlight of current status on finite-dimensional algebras and literature guide, cf.$\,$Remark~2.4;
 Arnav Tripathy
   for the topic course on 3d mirror symmetry/symplectic duality, fall 2016,
     that helped catching up many stringy frontier issues;
 Tristan Collins, Fabian Haiden, Sarah Harrison, Yang-Hui He, McLellan, Tripathy
   for communications, discussions, and/or literature guide on various themes at the frontier beyond the work;
 Collins, McLellan, Anand Patel, Shing-Tung Yau
   for other topic courses, fall 2016;
 Ling-Miao Chou
   for comments that improve the illustrations and moral support.
The project is supported by NSF grants DMS-9803347 and DMS-0074329.

\noindent ------------------------------------

 \noindent
 $^{\ast}$
 This version contains simplified {\sc Figure} 2-1 to limit the file size for the submission to arXiv.
 For complete illustrations, readers are referred to the version, posted one day behind,  in 							
  $\;$\verb=https://www.researchgate.net/profile/Chien-Hao_Liu/=$\;$
 %
 } 

\end{titlepage}

\newpage

\begin{titlepage}

$ $

\vspace{6em}

\centerline{\small\it
 Chien-Hao Liu dedicates this note to {\it Zhu Xiao-Mei}
  $($\raisebox{-.6ex}{\includegraphics{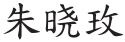}}$)$,}
\centerline{\small\it
 a pianist whose life story and performance on Bach touched his soul.}

\vspace{3em}
 
\centerline{\includegraphics[width=0.20\textwidth]{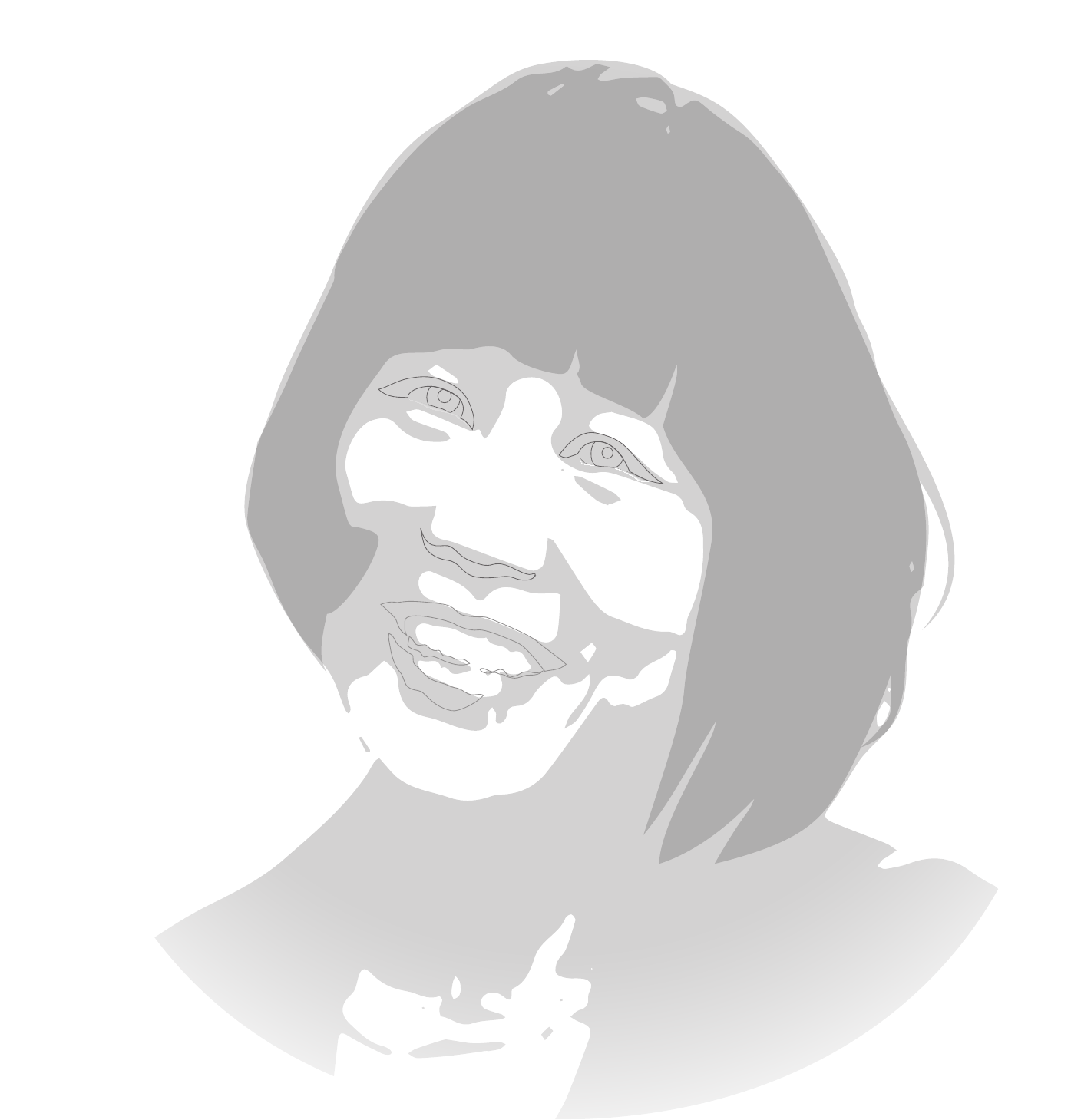}}

%
%
%

\end{titlepage}


\newpage
$ $

\vspace{-3em}

\centerline{\sc
 Admissible Condition on $(\varphi,\nabla)$
 } %

\vspace{2em}


\begin{flushleft}
{\Large\bf 0. Introduction and outline}
\end{flushleft}
In the previous work
  [L-Y] (D(13.1); arXiv:1606.08529 [hep-th]),
 we constructed the non-Abelian Dirac-Born-Infeld action $S_{DBI}$
      plus the Chern-Simons/Wess-Zumino term $S_{CS/WZ}$
 for D-branes as fundamental objects in string theory,
  whose world-volume are realized as differentiable maps
   $\varphi:(X^{\!A\!z},E; \nabla) \rightarrow Y$
   from an Azumaya/matrix manifold $X^{\!A\!z}$,
     with its defining fundamental module $E$ and a connection $\nabla$ on $E$,
   to a target-space $Y$ that is equipped with background fields $($metric, $B$-field, Ramond-Ramond field$)$
   from excitations of closed superstrings.
The first variation of the action and the corresponding equations of motion of $(\varphi,\nabla)$ were also derived ibidem.
In the process,
   on one hand as a requirement to resolve canonically a pull-push issue of tensors
     under a map from a noncommutative space (cf.\ $X^{\!A\!z}$) to a commutative space (cf.\ $Y$) and
   on the other hand as a requirement for $\nabla$ 	to be massless from the open-string aspect,
 an admissible condition was imposed on the pairs $(\varphi,\nabla)$
  ([L-Y: Definition 2.2.1, Remark 2.2.2, and Remark$\,$2.2.3] (D(13.1))).
 
The admissible condition ibidem consists of two parts: Condition (1) and Condition (2).
In this brief note, we examine these two conditions in more detail
  and bring their geometric implications on
    $(\varphi,\nabla)$ and the full action $S_{DBI}(\varphi,\nabla)+S_{CS/WZ}(\varphi,\nabla)$
	more transparent.
 In particular, we show that Condition (1)  alone already implies masslessness of $\nabla$ from open-string aspect;
  and that the additional Condition (2) implies a decoupling of the nilpotent fuzzy cloud of $\varphi(X^{\!A\!z})$
  to the dynamics of $(\varphi, \nabla)$.
We conclude with a refined definition of admissible $(\varphi,\nabla)$ and
   a remark on the anomaly factor in the integrand of the Chern-Simons/Wess-Zumino term $S_{CS/WZ}(\varphi,\nabla)$
   for D-branes based on the current study.

\bigskip
 
\bigskip

\noindent
{\bf Convention.}
 All the conventions and basic references follow
	  \begin{itemize}\small
	   \item[]  \hspace{-2em} [L-Y]\hspace{1em}\parbox[t]{40em}{{\it    	   	
        Dynamics of D-branes I.
          The non-Abelian Dirac-Born-Infeld action, its first variation, and the equations of motion for D-branes
		  --- with remarks on the non-Abelian Chern-Simons/Wess-Zumino term},
         arXiv:1606.08529 [hep-th]. (D(13.1))		 		
		 }
	  \end{itemize}  	

\bigskip

\bigskip
   
\begin{flushleft}
{\bf Outline}
\end{flushleft}
\nopagebreak
{\small
\baselineskip 12pt  
\begin{itemize}
 \item[0.]
  Introduction.
   
 \item[1.]
 The admissible condition on $(\varphi,\nabla)$
  in the construction of  the non-Abelian Dirac-Born-Infeld action $S_{DBI}(\varphi,\nabla)$
  \vspace{-.6ex}
  \begin{itemize}
   \item[\LARGE $\cdot$]
   Basic setup, terminology, and notations
   
   \item[\LARGE $\cdot$]
   The admissible condition on pairs $(\varphi,\nabla)$
  \end{itemize}
  
 \item[2.]
 Condition (1)
   \vspace{-.6ex}
   \begin{itemize}
    \item[\LARGE $\cdot$]
	 Generic uniformality of $(X_{\varphi},\,_{\varphi}{\cal E})$ over $X$
    
	\item[\LARGE $\cdot$]
	 Generic covariantly-invariant canonical splitting of $\varphi^{\sharp}$
	
    \item[\LARGE $\cdot$]
	 Condition (1) as a massless condition on $\nabla$ viewed from open strings in $Y$ via $\varphi$
   \end{itemize}
   
 \item[3.]
 Additional Condition (2)
  \vspace{-.6ex}
  \begin{itemize}
	\item[\LARGE $\cdot$]
	 Simplification of $S_{DBI}(\varphi,\nabla)+S_{CS/WZ}(\varphi,\nabla)$

    \item[\LARGE $\cdot$]	
	 A refinement of admissibility
	
	\item[\LARGE $\cdot$]
	 The anomaly factor in the Chern-Simons/Wess-Zumino term $S_{CS/WZ}(\varphi,\nabla)$
  \end{itemize}
 %
 %
 %
 
\end{itemize}
} 

\newpage

\section{The admissible condition on $(\varphi,\nabla)$
  in the construction of  the non-Abelian Dirac-Born-Infeld action $S_{DBI}(\varphi,\nabla)$}

\begin{flushleft}
{\bf Basic setup, terminology, and notations}
\end{flushleft}
Given
  a (real, smooth) manifold $X$ and
  a complex (smooth) vector bundle with a (smooth) connection $(E,\nabla)$ over $X$,
 let
  $$
     X^{\!A\!z}\;=\; (X, C^{\infty}(\End_{\Bbb C}(E)))
  $$
  be the {\it noncommutative manifold}
    with the underlying topology $X$  but
	with the function ring the noncommutative ring $C^{\infty}(\End_{\Bbb C}(E))$
	of smooth endomorphisms of $E$.
 Let $Y$ be another smooth manifold.	
 A {\it smooth map} (i.e.\ $C^{\infty}$-map)
  $$
   \varphi\;:\; (X^{\!A\!z},E;\nabla)\;\longrightarrow\; Y
  $$
  is {\it defined contravariantly by} a ring-homomorphism
  $$
     C^{\infty}(\End_{\Bbb C}(E))\;\longleftarrow\; C^{\infty}(Y)\;:\; \varphi^{\sharp}
  $$
  over the built-in inclusion ${\Bbb R}\subset {\Bbb C}$.

$\varphi^{\sharp}$ extends canonically to a commutative diagram of ring-homomorphisms
 (over ${\Bbb R}$ or ${\Bbb R}\subset {\Bbb C}$, whichever is applicable):
 $$
   \xymatrix{
	 C^{\infty}(\End_{\Bbb C}(E))
                 	&&& C^{\infty}(Y)\ar[lll]_-{\varphi^{\sharp}}
								                           \ar[llld]^-{f_{\varphi}^{\sharp}}
														   \ar@{_{(}->}[dd]^-{pr_Y^{\sharp}}	 \\
         A_{\varphi}\rule{0ex}{1em} \ar@{^{(}->}[u]_-{\sigma_{\varphi}^{\sharp}}   \\
      \;\;C^{\infty}(X)\;\; \rule{0ex}{3ex}  \ar@{^{(}->}[u]_-{\pi_{\varphi}^{\sharp}}
																	 \ar@{^{(}->}[rrr]_-{pr_X^{\sharp}}
         &&&   C^{\infty}(X\times Y)\ar@{->>}[ulll]_-{\tilde{f}_{\varphi}^{\sharp}}
   }
  $$
  where
    $\pr_X:X\times Y \rightarrow X$ and $\pr_Y:X\times Y\rightarrow Y$ are the projection maps.
 
{\it Which defines in turn} a commutative diagram of smooth maps between the associated noncommutative manifold,
 manifold, or $C^{\infty}$-scheme:
  $$
   \xymatrix{
     X^{\!A\!z}\ar[rrrr]^-{\varphi}\ar@{->>}[d]^-{\sigma_{\varphi}}  &&&& Y \\
     \;\;X_{\varphi}\ar[rrrru]_-{f_{\varphi}}\;\;
	                                   \ar@{_{(}->} [rrrrd]^-{\tilde{f}_{\varphi}}
                                	   \ar@{->>}[d]^-{\pi_{\varphi}}          \\
	 X     &&&& X\times Y \ar@{->>}[uu]_-{pr_Y} \ar@{->>}[llll]^-{pr_X}
    }
  $$
The $C^{\infty}$-scheme $X_{\varphi}$ is called the {\it surrogate}  of $X^{\!A\!z}$ specified by $\varphi$.
  
In many situations, it is more convenient to proceed in terms of sheaves and stalks.
For that we denote
 ${\cal O}_X$ the structure sheaf of smooth functions on $X$,
 ${\cal O}_X^{\,\Bbb C}:={\cal O}_X\otimes_{\Bbb R}{\Bbb C}$ its complexification,
 ${\cal E}$ the sheaf of smooth sections of $E$,
 $\Endsheaf_{{\cal O}_X^{\,\Bbb C}}({\cal E})$ the sheaf of endomorphisms algebras of $E$,
 ${\cal A}_{\varphi}={\cal O}_{X_{\varphi}}$ the structure sheaf of (smooth functions on) $X_{\varphi}$.
As the fundamental $\Endsheaf_{{\cal O}_X^{\,\Bbb C}}({\cal E})$-module,
  ${\cal E}$ is naturally an ${\cal O}_{X_{\varphi}}^{\,\Bbb C}$-module on $X_{\varphi}$,
  which will be denoted by $_{\varphi}{\cal E}$.
  
See [L-Y: Sec.\ 2.1] for a more detailed concise review.

\bigskip

\begin{flushleft}
{\bf The admissible condition on pairs $(\varphi,\nabla)$}
\end{flushleft}
Recall  the following admissible condition on $(\varphi,\nabla)$:
   ([L-Y: Definition 2.2.1] (D(13.1))
 
\bigskip

\begin{sdefinition} {\bf [admissible pair $(\varphi,\nabla)$]}$\;$ {\rm
 Let $\varphi:(X^{\!A\!z},E)\rightarrow Y$ be a differentiable map  and
  $\nabla$ be a connection on $E$.
 The pair $(\varphi,\nabla)$ is called {\it admissible} if the following two conditions are satisfied
  over an open-dense subset of $X$:
   $$
    (1)\;\;\;
     DA_{\varphi}\;\subset\; C^{\infty}(\Omega_X)\otimes_{C^{\infty}(X)}A_{\varphi}
	  \hspace{2em}\mbox{and}\hspace{2em}
	(2)\;\;\;
	 F_{\nabla}\;
	    \subset\;  C^{\infty}(\Omega_X^2)\otimes_{C^{\infty}(X)} \Comm(A_{\varphi})\,.
   $$
 Here,
  $\Comm(A_{\varphi})$ is the commutant of $A_{\varphi}$
     in $C^{\infty}(\End_{\Bbb C}(E))$.
 For convenience, we say also that
  $\varphi$ is an {\it admissible map} from $(X^{\!A\!z},E;\nabla)$ to $Y$,   or that
  $\varphi:(X^{\!A\!z},E)\rightarrow Y$ is a map that is {\it admissible} to $\nabla$ on $E$,   or that
  $\nabla$ is a connection on $E$ that is {\it admissible} to $\varphi:(X^{\!A\!z},E)\rightarrow Y$.
}\end{sdefinition}

\bigskip

Condition (1) allows
  the pull-back $f_{\varphi}^{\sharp}(\,\tinybullet\,)$ of a tensor $\tinybullet$ on $Y$
    to the surrogate $X_{\varphi}$
 to descend canonically to an $\End_{\Bbb C}(E)$-valued tensor
  $\varphi^{\diamond}(\,\mbox{\LARGE $\cdot$}\,)
     := \pi_{\varphi,\ast} f_{\varphi}^{\ast}  (\,\mbox{\LARGE $\cdot$}\,)    $
  on $X$.
The latter can then be combined with the curvature tensor $F_{\nabla}$ of $\nabla$ on $X$ to
  define the term
  $$
    \varphi^{\ast}(g+B)\:+\: 2\pi\alpha^{\prime}\,F_{\nabla}\;
	 :=\;    \varphi^{\diamond}(g+B)\:+\: 2\pi\alpha^{\prime}\,F_{\nabla}
  $$
  in the Dirac-Born-Infeld action $S_{\DBI}$ for $(\varphi,\nabla)$ rigorously.
Condition (2) is a massless-from-open-string-viewpoint condition on $\nabla$.
See [L-Y: Sec.\ 2.2 \& Sec.\ 3.1.1] (D(13.1)) for details.
(See also Sec.$\,$2 of the current work for an improved understanding of Condition (1),
    which says that it alone already implies masslessness of $\nabla$; cf.\ Corollary~2.11.)
\bigskip

\begin{sremark} $[$Condition (1) $\Rightarrow$ weaker Condition (2$^{\,\prime}$)$\,]\;$ {\rm
 Observe that
 the curvature $F_D :=[D,D]$ of the induced connection $D$ on $\End_{\Bbb C}(E)$
  from the connection $\nabla$ on $E$
  can be expressed in terms of the curvature $F_{\nabla}$ of $\nabla$ as
  $F_D =[F_{\nabla}\,,\,\cdot\;]$,
 an identity that follows from a straightforward computation.
 (However, see [DV-M: Sec.\ 2: near the end] for a related discussion in the bigger context
     of noncommutative differential geometry of the ring $C^{\infty}(\End_{\Bbb C}(E))$.)
 It follows that
  if Condition (1) holds,
   then
     $$
	  (2^{\prime})\hspace{6em}
	   F_{\nabla}\subset
         C^{\infty}(\Omega^2_X)\otimes_{C^{\infty}(X)}\Comm^{(1)}(A_{\varphi})\,,
		\hspace{7em}
	  $$
  where $\Comm^{(1)}(A_{\varphi})$ consists of all elements $s\in C^{\infty}(\End_{\Bbb C}(E))$
    such that $[s,A_{\varphi}]\subset A_{\varphi}$.
}\end{sremark}

\bigskip

\noindent
{\it Convention and Terminology}$\;$ {\rm
 If needed, we'll denote the `{\sl open dense subset of $X$}' in Definition~1.1 by $U$.
 A property that holds over $U$ (but may not hold over the whole $X$) is called {\it generic over $X$}.
 For the simplicity of presentation, we may proceed throughout this work as if $U=X$
   or use the word `{\it generic}' to mean `{\it defined or valid only over $U\subsetneqq X$}'.
} 

\bigskip

\section{Condition (1)}

Let $(\varphi,\nabla)$ be a pair that satisfies Condition (1)
  $$
      DA_{\varphi}\;\subset\; C^{\infty}(\Omega_X)\otimes_{C^{\infty}(X)}A_{\varphi}
  $$
 generically.
We show in this section
 that this implies a generically uniform geometry of $(X_{\varphi},\,_{\varphi}{\cal E})$ over $X$   and
 that the condition alone can already serve as a massless condition of $\nabla$
  viewed from the open strings on $Y$ via $\varphi$.

\bigskip

\begin{flushleft}
{\bf Generic uniformality of $(X_{\varphi},\,  _{\varphi}{\cal E})$ over $X$}
\end{flushleft}

\begin{slemma} {\bf [generic uniformality of $X_{\varphi}$ over $X$]}$\;$
 The built-in ${\cal O}_X$-module inclusion
   ${\cal A}_{\varphi}\subset \Endsheaf_{{\cal O}_X^{\,\Bbb C}}({\cal E})$
   is indeed a $D$-invariant ${\Bbb R}$-subalgebra bundle inclusion
   $\mbox{\boldmath $A$}_{\varphi}\subset \End_{\Bbb C}(E)$ over $X$.
 Furthermore, all fibers of the bundle $\mbox{\boldmath $A$}_{\varphi}$
  (which are canonically identified with the  corresponding fibers of the ${\cal O}_X$-module ${\cal A}_{\varphi}$)
  are isomorphic ${\Bbb R}$-algebras.
\end{slemma}

\smallskip

\begin{proof}
 Since $D$ is a linear connection on a vector bundle,
  Condition (1) can be rephrased in terms of $D$-parallel transports as
   \begin{itemize}
    \item[\LARGE $\cdot$] {\it
	 Given a section $s\in A_{\varphi}\subset C^{\infty}(\End_{\Bbb C}(E))$
	    and any vector field $\xi$ on $X$,
     the $D$-parallel transport of $s$ along the flow on $X$ generated by $\xi$ must remain in $A_{\varphi}$.}
  \end{itemize}
 Since $X$ is smooth (and connected), any two points can be connected by a finite integral trajectory of some $\xi$.
 Note also that $D$-parallel transports are invertible.
 It follows that
  $D$-parallel transports take
   stalk of ${\cal A}_{\varphi}$ to stalks of ${\cal A}_{\varphi}$
    isomorphically as substalks of stalks of $\Endsheaf_{{\cal O}_X^{\,\Bbb C}}({\cal E})$ and hence
   fibers of ${\cal A}_{\varphi}$ to fibers of ${\cal A}_{\varphi}$
    isomorphically as subfibers of fibers of $\Endsheaf_{{\cal O}_X^{\,\Bbb C}}({\cal E})$.
 This shows that
   the built-in ${\cal O}_X$-module inclusion
   ${\cal A}_{\varphi}\subset \Endsheaf_{{\cal O}_X^{\,\Bbb C}}({\cal E})$
   is indeed a $D$-invariant ${\Bbb R}$-subbundle inclusion
   $\mbox{\boldmath $A$}_{\varphi}\subset \End_{\Bbb C}(E)$ over $X$.
 Since ${\cal A}_{\varphi}$ is a sub-${\cal O}_X$-algebra of
   $\Endsheaf_{{\cal O}_X^{\,\Bbb C}}({\cal E})$,
 the subbundle inclusion $\mbox{\boldmath $A$}_{\varphi}\subset \End_{\Bbb C}(E)$
   is in addition an ${\Bbb R}$-subalgebra bundle inclusion.

 To see all the fibers of $\mbox{\boldmath $A$}_{\varphi}$  are isomorphic ${\Bbb R}$-algebras,
  note that
  \begin{itemize}
   \item[\LARGE $\cdot$] {\it
   If $D_{\xi}s_1=D_{\xi}s_2=0$, then
    $D_{\xi}(s_1\cdot s_2)= (D_{\xi}s_1)\cdot s_2+ s_1\cdot D_{\xi}s_2  =0$,\\
	where $\cdot$ is the multiplication of sections in $\End_{\Bbb C}(E)$.}
  \end{itemize}
  It follows that $D$-parallel transports are ${\Bbb R}$-algebra-homomorphisms.
  Since  parallel transports are invertible, they are isomorphisms.
 This proves the lemma.
  
\end{proof}

\bigskip

\begin{slemma} {\bf [generic uniformality of $_{\varphi}{\cal E}$ over $X$]}$\;$	
 Continuing Lemma~2.1.
 Denote this finite-dimensional ${\Bbb R}$-algebra by $R$ and
	 let $R^{\,\Bbb C}:= R\otimes_{\Bbb R}{\Bbb C}$ be its complexification.
 Then,
  all $(_{\varphi}{\cal E})|_x :=  (_{\varphi}{\cal E})|_{\pi_{\varphi}^{-1}(x)}$, $x\in X$,
  are isomorphic as $R^{\,\Bbb C}$-modules.
\end{slemma}

\smallskip

\begin{proof}
 This follows from the fact that
  \begin{itemize}
   \item[\LARGE $\cdot$]{\it
   Let $s\in C^{\infty}(\End_{\Bbb C}(E))$ and $v\in C^{\infty}(E)$
    such that $D_{\xi}s=0$ and $\nabla_{\xi}v=0$.\\
  Then 	
    $\nabla_{\xi}(sv)= (D_{\xi}s)v + s\, \nabla_{\xi}v=0$.}
  \end{itemize}
\end{proof}

\noindent
Cf.\ {\sc Figure}~2-1.
%

\begin{figure}[htbp]
 \bigskip
  \centering
   \includegraphics[width=0.80\textwidth]{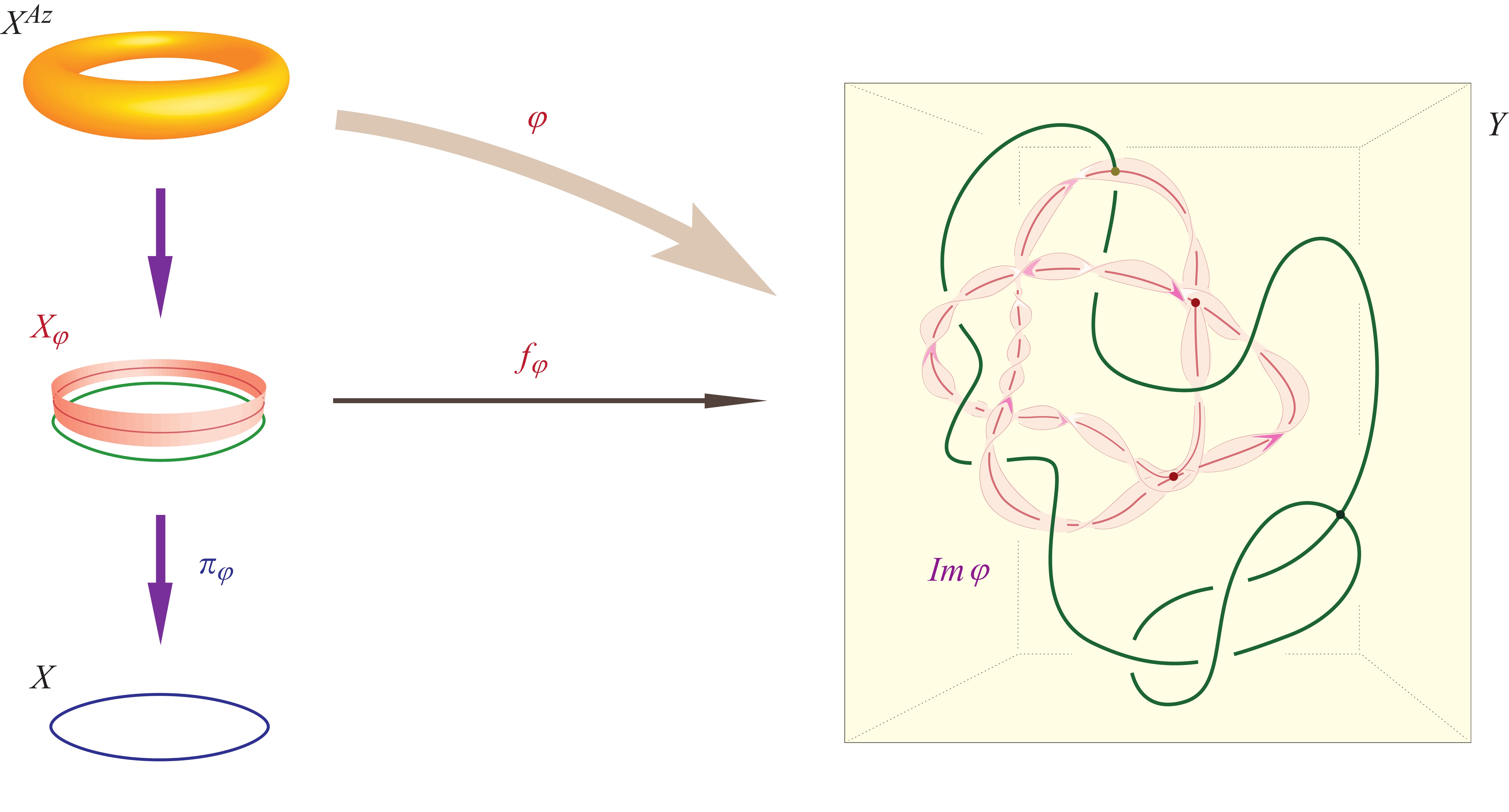}

  \bigskip
  
  \bigskip
  
 \centerline{\parbox{13cm}{\small\baselineskip 12pt
  {\sc Figure}~2-1.
  A map $\varphi:(X^{\!A\!z},E)\rightarrow Y$ that has its specified surrogate $X_{\varphi}$ unifrom over $X$ is indicated.
  Note that the image $\varphi(X^{\!A\!z})$ of $\varphi$ in $Y$ can still have very rich variations of behavior
   as a $C^{\infty}$-subscheme of $Y$.
  When $(\varphi,\nabla)$ satisfies Condition$\:$(1),
   both $X_{\varphi}$ and the induced sheaf $_{\varphi}{\cal E}$ on $X_{\varphi}$
   are generically uniform over $X$.
 In the particular example illustrated, $X_{\varphi}$ has two connected components, one (green) is reduced
  and the other (red) is nonreduced. The nilpotent fuzzy structure of the latter is indicated by a pink ribbon.
 The image {\it Im}$\,\varphi$ of $\varphi$ in $Y$ is thus a non-reduced $C^{\infty}$-subscheme of $Y$.
 Its generic nilpotent fuzzy cloud is indicated by a rotating pink ribbon in $Y$.
 There can be distinct $x$, $x^{\prime }\in X$
   such that $f_{\varphi}(\pi_{\varphi}^{-1}(x))$ and  $f_{\varphi}(\pi_{}^{-1}(x^{\prime}))$
    meet in $Y$.
 At such an intersection point $p$ in $Y$, the non-reduced structure of {\it Im}$\,\varphi$ can be enhanced, depending on
  how $f_{\varphi}(\pi_{\varphi}^{-1}(U_x))$ and
          $f_{\varphi}(\pi_{\varphi}^{-1}(U_{x^{\prime}}))$
    meet in $Y$ at $p$.
 Here, $U_x$, $U_{x^{\prime}}$ are small neighborhoods of $x$, $x^{\prime}\in X$ respectively.
  }}
\end{figure}

\bigskip

The generic uniformality of $X_{\varphi}$ over $X$ implies immediately that:

\bigskip

\begin{scorollary} {\bf [$(X_{\varphi})_{\redscriptsize}$ generic covering space of $X$]}$\;$
 Let
  $(X_{\varphi})_{\redscriptsize}$ be the reduced subscheme of $X_{\varphi}$.
 Then $(X_{\varphi})_{\redscriptsize}$ is a manifold generically and
  the built-in map $\pi_{\varphi}: X_{\varphi}\rightarrow X$ restricts to a generic covering map
  $$
    \pi_{\varphi}|_{(X_{\varphi})_{\redtiny}}\;:\;
	 (X_{\varphi})_{\redscriptsize}\; \longrightarrow\; X\,.
  $$
\end{scorollary}

\bigskip
 
\begin{sremark} $[$finite-dimensional algebra and module$]\;$ {\rm
 Note that the moduli of isomorphism classes
  of finite-dimensional algebras or  finite dimensional modules of a given finite-dimensional algebra
  can be a continuum.
 See, for example, [Ca], [Je1], [Je2], [Poo1], [Poo2] for details.
 It follows that though $(X_{\varphi},\,_{\varphi}{\cal E})$ is uniform over $X$,
  the structure of $(X_{\varphi},\,_{\varphi}{\cal E})$ remains very rich.
}\end{sremark}

\bigskip

\begin{snotation} {\bf [idempotent section of ${\cal A}_{\varphi}$]}$\;$ {\rm
 For $x\in X$, denote by $q\in \pi_{\varphi}^{-1}(x)$ an ${\Bbb R}$-point
  in the $0$-dimensional $C^{\infty}$-subscheme $\pi_{\varphi}^{-1}(x)$  of $X_{\varphi}$.
 The number of $q$'s  is the same as the generic degree of the generic covering map
   $\pi_{\varphi}|_{(X_{\varphi})_{\redtiny}}: (X_{\varphi})_{\redscriptsize}\rightarrow X$.
  As a sheaf of  ${\cal O}_X$-algebras,
   the stalk $({\cal A}_{\varphi})_x$ of ${\cal A}_{\varphi}$
    is canonically decomposed into a product of ${\cal O}_{X,\,x}$-algebras
   $$
     ({\cal A}_{\varphi})_x\;
	   =\; \times_{q\in\pi_{\varphi}^{-1}(x)}\,({\cal A}_{\varphi})_x\,e_q\,,
   $$
    where $e_q\in ({\cal A}_{\varphi})_x$ are idempotent:
	$e_q^2=e_q$ and $\sum_{q\in \pi_{\varphi}^{-1}(x) }e_q=1$.
 Such $e_q$'s are unique and are $D$-covariantly constant: $D_{\mbox{\tiny $\bullet$}}\,e_q=0$
  (since
      $D_{\mbox{\tiny $\bullet$}}\,e_q=D_{\mbox{\tiny $\bullet$}}\,e_q^2
	     = 2\, e_q\,D_{\mbox{\tiny $\bullet$}}\,e_q$, which implies that
	  $D_{\mbox{\tiny $\bullet$}}\,e_q=  e_q\,D_{\mbox{\tiny $\bullet$}}e_q
	     =2\,e_q\,D_{\mbox{\tiny $\bullet$}}e_q$ and, hence, must be zero).
}\end{snotation}

\bigskip

\begin{flushleft}
{\bf Generic covariantly-invariant canonical splitting of $\varphi^{\sharp}$}
\end{flushleft}
\begin{slemma}
 {\bf [generic canonical splitting of ${\cal O}_{X_{\varphi}}$ and $\varphi^{\sharp}$]}$\;$
 Take $X=U$.
 $(1)$
 Let ${\cal I}_{(X_{\varphi})_{\redtiny}}$ be the ideal sheaf of
   $(X_{\varphi})_{\redscriptsize}$ in $X_{\varphi}$.
 Then, the exact sequence of  ${\cal O}_{X_{\varphi}}$-modules
   $$
    0\; \longrightarrow\;
         {\cal I}_{(X_{\varphi})_{\redtiny}} \;
         \longrightarrow\;  {\cal O}_{X_{\varphi}}\;
         \longrightarrow\;  {\cal O}_{(X_{\varphi})_{\redtiny}}\; \longrightarrow\; 0
  $$
  splits by a canonical ${\cal O}_{X_{\varphi}}$-algebra inclusion
  $$
   \varpi^{\varphi,\,\sharp}_{\redscriptsize}\;:\; {\cal O}_{(X_{\varphi})_{\redtiny}}\;
      \hookrightarrow\;   {\cal O}_{X_{\varphi}}\,.
  $$
 Identify ${\cal O}_{(X_{\varphi})_{\redtiny}}$
  as an ${\cal O}_X$-submodule of $\Endsheaf_{{\cal O}_X^{\,\Bbb C}}({\cal E})$
  via $\varpi_{\redscriptsize}^{\varphi,\,\sharp}$,
 then
  $$
    {\cal O}_{X_{\varphi}}\;
	     =\;  {\cal O}_{(X_{\varphi})_{\redtiny}}\,+\, {\cal I}_{(X_{\varphi})_{\redtiny}}
  $$
  in $\Endsheaf_{{\cal O}_X^{\,\Bbb C}}({\cal E})$.
  
 $(2)$
 As ${\cal O}_X$-submodules of $\Endsheaf_{{\cal O}_X^{\,\Bbb C}}({\cal E})$,
   both ${\cal O}_{(X_{\varphi})_{\redtiny}}$
     and ${\cal I}_{(X_{\varphi})_{\redtiny}}$
   are covariantly invariant under the connection $D$ on $\Endsheaf_{{\cal O}_X^{\,\Bbb C}}({\cal E})$.
 In particular, given a local section
   $s\in {\cal O}_{X_{\varphi}}\subset \Endsheaf_{{\cal O}_X^{\,\Bbb C}}({\cal  E})$,
  let $s=s_0+n_s$ be the decomposition of $s$ according to Statement $(1)$.
 Then,
  $$
    D_{\mbox{\tiny $\bullet$}\,}s\;
	 =\;    D_{\mbox{\tiny $\bullet$}\,}s_0\; +\;  D_{\mbox{\tiny $\bullet$}\,}n_s\;
  $$
   is the decomposition of $D_{\mbox{\tiny $\bullet$}\,}s$ according to Statement $(1)$.
	
 $(3)$
 The ring-homomorphism
  $\varphi^{\sharp}:C^{\infty}(Y)\rightarrow C^{\infty}(\End_{\Bbb C}(E))$
  decomposes accordingly (and hence canonically) into
  $$
    \varphi^{\sharp}\;=\; \varphi^{\sharp}_{\redscriptsize}\;+ \xi_{\varphi}\,,
  $$
 where
    $$
	  \varphi^{\sharp}_{\redscriptsize}\;:\;
	   C^{\infty}(Y)\;\longrightarrow\; C^{\infty}(\End_{\Bbb C}(E))
	$$
	  is a ring-homomorphism over ${\Bbb R}\hookrightarrow{\Bbb C}$
	  whose surrogate $X_{\varphi_{\redtiny}}$ is $(X_{\varphi})_{\redscriptsize}$, and
    $$
	  \xi_{\varphi}\;:\; C^{\infty}(Y)\;\longrightarrow\; C^{\infty}(\End_{\Bbb C}(E))
	$$
 	 is a derivation on $C^{\infty}(Y)$ taking values in the nilpotent elements of
	 $A_{\varphi}\subset C^{\infty}(\End_{\Bbb C}(E))$.
 $\;\xi_{\varphi}$ corresponds to the nilpotent fuzziness of the image brane $\varphi(X^{\!A\!z})$ in $Y$.
\end{slemma}

\smallskip

\begin{proof}
 Since the built-in map $(X_{\varphi})_{\redscriptsize}\rightarrow X$ is a covering map,
  the built-in diagram of maps
   $$
     \xymatrix{
      \hspace{1em}X_{\varphi} \ar@{->>}[rd]_-{\pi_{\varphi}}\hspace{1em}
	    &&  (X_{\varphi})_{\redscriptsize}  \ar@{->>}[ld] \\
	       & X
       }	
   $$
  extends uniquely to a commutative diagram of maps
   $$
     \xymatrix{
      \hspace{1em}X_{\varphi} \ar@{->>}[rd]_-{\pi_{\varphi}}\hspace{1em}
	     \ar@{->>}[rr]^-{\varpi^{\varphi}_{\redtiny}}
	    &&  (X_{\varphi})_{\redscriptsize}  \ar@{->>}[ld] \\
	       & X
       }	
   $$
   such that the composition
   $$
    \xymatrix{
     (X_{\varphi})_{\redscriptsize}\hspace{1ex}  \ar@{^{(}->}[r]
	    &  \hspace{1ex}X_{\varphi}\hspace{1ex} \ar@{->>}[r]^-{\varpi^{\varphi}_{\redtiny}}
		& \hspace{1ex}(X_{\varphi})_{\redscriptsize}
	}
   $$
   is the identity map on $(X_{\varphi})_{\redscriptsize}$.
 This proves Statement (1).
  
 Recall Notation~2.5.
 Then, for the stalk at $x\in X$,
   the identification of ${\cal O}_{(X_{\varphi})_{\redtiny}}$
     as an ${\cal O}_X$-submodule of $\Endsheaf_{{\cal O}_X^{\,\Bbb C}}({\cal E})$
	 via $\varpi_{\redscriptsize}^{\varphi,\,\sharp}$
    is simply $\oplus_{q\in\pi_{\varphi}^{-1}(x)}{\cal O}_{X,x}e_q$.	
 Since ${\cal O}_{X,x}$ is covariantly invariant and $e_q$'s are covariantly constant under the connection $D$,
  ${\cal O}_{(X_{\varphi})_{\redtiny}}$ is covariantly invariant under $D$.	  	
 Since
   parallel transports from the connection $D$ on $\End_{\Bbb C}(E)$
     are algebra-isomorphisms on fibers of $\End_{\Bbb C}(E)$    and
   Condition (1) implies that they leaves $A_{\varphi}$ and, hence, ${\cal O}_{X_{\varphi}}$ invariant,
  they must leave the nilradical
    ${\cal I}_{(X_{\varphi})_{\redtiny}}$ of ${\cal O}_{X_{\varphi}}$ invariant.
 This proves Statement (2).
  
 Recall that $C^{\infty}(X_{\varphi})=A_{\varphi}$.
 The composition
  $$
   \xymatrix{
     C^{\infty}(Y)\;   \ar[r]^-{\varphi^{\sharp}}
	 & \; A_{\varphi}\;  \ar@{->>}[r]
   	 & \; C^{\infty}((X_{\varphi})_{\redscriptsize})\;\;
	      \ar@{^{(}->}[r]^-{\varpi^{\varphi,\sharp}_{\redtiny}}
     & \; A_{\varphi}\; \ar@{^{(}->}[r]
	 & \;C^{\infty}(\End_{\Bbb C}(E))
    }		
  $$	
  defines a ring-homomorphism
  $$
     \varphi_{\redscriptsize}^{\sharp}\;:\; C^{\infty}(Y)\;
	    \longrightarrow\;  C^{\infty}(\End_{\Bbb C}(E))
  $$
  over ${\Bbb R}\hookrightarrow{\Bbb C}$
  with $A_{\varphi_{\redtiny}}=C^{\infty}((X_{\varphi})_{\redscriptsize})
             \subset A_{\varphi}$.
 Let
  $$
     \xi_{\varphi}\,:=\, \varphi^{\sharp}-\varphi_{\redscriptsize}^{\sharp}\;:\;
	   C^{\infty}(Y)\; \longrightarrow\; C^{\infty}(\End_{\Bbb C}(E))\,.
  $$
 Then, by construction,
   $\xi_{\varphi}$ is a derivation on $C^{\infty}(Y)$ taking values in the nilpotent elements of
	 $A_{\varphi}\subset C^{\infty}(\End_{\Bbb C}(E))$.
 Since $\varphi^{\sharp}_{\redscriptsize}$ is canonical, so is $\xi_{\varphi}$.
 This proves Statement (3).
   
\end{proof}

\bigskip

\begin{sdefinition}
{\bf [reduced component and nilpotent component of $\varphi^{\sharp}$]}$\;$ {\rm
 We will call
  $\varphi^{\sharp}_{\redscriptsize}$ in the decomposition
     the {\it reduced component} of $\varphi^{\sharp}$ and
  $\xi_{\varphi}$ the {\it nilpotent component} of $\varphi^{\sharp}$.
 Caution that the decomposition and the resulting components are in general defined only over $U$.
}\end{sdefinition}

\bigskip

The ring-homomorphism $\varphi^{\sharp}_{\redscriptsize}$ over ${\Bbb R}\hookrightarrow{\Bbb C}$
 defines a smooth map (over $U$)
  $$
    \varphi_{\redscriptsize}\; :\;   (X^{\!A\!z},E;\nabla)\;  \longrightarrow\; Y
  $$
 and
 one has the following commutative diagram that relates
  all the built-in objects and maps underlying $\varphi$ and $\varphi_{\redscriptsize}$:
 $$
   \xymatrix{
    X^{\!A\!z}   \ar@<+.4ex>[rrrrr]^-{\varphi} \ar@{=}[rd]   \ar@{->>}[d]_-{\sigma_{\varphi}} &&&&& Y \\
	X_{\varphi}  \ar@/^2ex/[rrrrru]^(.35){f_{\varphi}}
	                                |(.22){\mbox{\rule{0ex}{1ex}$\hspace{3em}$}}
								  \ar@{->>}[rdd]^-{\varpi^{\varphi}_{\redtiny}}	
								  \ar@<-.6ex>@{->>}[rddd]_-{\pi_{\varphi}}
	    &   X^{\!A\!z}\ar[rrrru]_-{\rule{1em}{0ex}\varphi_{\redtiny}}
		                             \ar@{->>}[dd]^-{\sigma_{\varphi_{\redtiny}}}  \\ \\
	    &  \rule{0ex}{1.4em}\hspace{16ex}
		     X_{\varphi_{\redtiny}}=\;(X_{\varphi})_{\redtiny}\hspace{6ex}
		            \ar@/^1ex/[rrrruuu]_-{\hspace{3ex}f_{\varphi_{\redtiny}} =\; (f_{\varphi})_{\redtiny}}
                     \ar@{->>}[d]^-{\pi_{\varphi_{\redtiny}}}	
					 \ar@<+.6ex>@{^{(}->}[luu]	\\			
		& X
   }
 $$	
 over the open dense subset $U$ of $X$.
 
%

\bigskip

\begin{flushleft}
{\bf Condition (1) as a massless condition on $\nabla$ viewed from open strings in $Y$ via $\varphi$}
\end{flushleft}
\begin{slemma} {\bf [induced connection on $_{\varphi_{\redtiny}}\!{\cal E}$]}$\;$
 Take $X=U$.
 $(1)$
  The ${\cal O}_{X_{\varphi_{\redtiny}}}^{\,\Bbb C}$-module
   $_{\varphi_{\redtiny}}\!{\cal E}$ is locally free; i.e. a sheaf of sections of a complex vector bundle
  $_{\varphi_{\redtiny}}\!E$.
 $\;\;(2)$
 The connection $\nabla$ on $E$ induces
   a connection $_{\varphi_{\redtiny}}\!\nabla$ on $_{\varphi_{\redtiny}}\!E$,
   (equivalently on $_{\varphi_{\redtiny}}\!{\cal E}$ ).
\end{slemma}
 
\smallskip

\begin{proof}
 Statement (1)  is an immediate consequence of Lemma~2.2.
 Consider now Statement (2).

 Since $(\pi_{\varphi_{\redtiny}})_{\ast}(_{\varphi_{\redtiny}}{\cal E})$
   is canonically isomorphic to ${\cal E}$,
 for any $x\in X$ the stalk ${\cal E}_x$ is canonically isomorphic to  the stalk
  $\left((\pi_{\varphi_{\redtiny}})_{\ast}(_{\varphi_{\redtiny}}{\cal E})\right)_x$,
  which in turn is canonically isomorphic to
  $\oplus_{q\in \pi_{\varphi_{\redtiny}}^{-1}(x)}(_{\varphi_{\redtiny}}\!{\cal E})_q$
  since $\pi_{\varphi_{\redtiny}}:X_{\varphi_{\redtiny}}\rightarrow X$ is a covering map.
 
 On the other hand,
  recall the covariantly constant idempotent elements $e_q$, $q\in \pi_{\varphi}^{-1}(x)$,
     from Notation~2.5.
 Then,
  $$
	 {\cal E}_x\;=\; \oplus_{q\in\pi_{\varphi}^{-1}}\,e_q\,{\cal E}_x\,.
  $$
 Furthermore,
  let $\xi\in Der_{\Bbb R}({\cal O}_{X,x})$ be a germ of derivations and $v\in e_q{\cal E}_x$;
 then
  $$
    \nabla_{\xi}v\;=\; \nabla_{\xi}(e_qv)\;=\; (D_{\xi}e_q)v\,+\, e_q\,\nabla_{\xi}v\;
	 =\; e_q\,\nabla_{\xi}v\;\in e_q\,{\cal E}_x\,.
  $$
 It follows that
  \begin{itemize}
   \item[\LARGE $\cdot$] {\it
   The decomposition  $\; {\cal E}_x = \oplus_{q\in\pi_{\varphi}^{-1}}\,e_q\,{\cal E}_x\;$ above
    is $\nabla$-invariant.}
  \end{itemize}
    
 With the canonical identification of
  a neighborhood of $x\in X$ with a neighborhood of $q\in X_{\varphi_{\redtiny}}$,
    via $\pi_{\varphi_{\redtiny}}$,
   and
  $e_q\,{\cal E}_x$ with $(_{\varphi_{\redtiny}}{\cal E})_q$,
   via $(\pi_{\varphi_{\redtiny}})_{\ast}$,
 Statement (2) follows.
    
\end{proof}

\bigskip

The proof above gives indeed:

\bigskip

\begin{slemma} {\bf [decomposition of $\nabla$ on stalks of ${\cal E}$ that matches $\varphi$]}$\;$
 Denote  by $^q\nabla$
   the induced connection from $\nabla$ on the $\nabla$-invariant direct summand $e_q\,{\cal E}_x$.
 Then,
  $$
    ({\cal E}_x,\,\nabla)\;
	   =\; \oplus_{q\in\pi_{\varphi}^{-1}}\,(e_q\,{\cal E}_x,\,^q\nabla)\,.
  $$
\end{slemma}

\bigskip

Note that,, taking $X=U$, the complex vector bundle $_{\varphi_{\redtiny}}\!E$ on $X_{\varphi_{\redtiny}}$
  in general has different rank over different connected components of $X_{\varphi_{\redtiny}}$.

The following corollary is an immediate consequence of Lemma~2.9:
  
\bigskip

\begin{scorollary} {\bf [Condition ({\boldmath $2^{\prime\prime}$})]}$\;$
 Condition (1) implies another condition weaker than but resembling Condition (2):
   $$
	  (2^{\prime\prime})\hspace{6em}
	   F_{\nabla}\subset
         C^{\infty}(\Omega^2_X)
		   \otimes_{C^{\infty}(X)}\Comm((A_{\varphi})_{\redscriptsize})\,.
		\hspace{7em}
	  $$
  Here,
      $(A_{\varphi})_{\redscriptsize}
	      = \Gamma({\cal A}_{\varphi}/\mbox{\it ${\cal N}\!$il${\cal R}$adical}\,({\cal A}_{\varphi}))$,
	   with $\Gamma$ the global smooth section functor on ${\cal O}_X$-modules.
\end{scorollary}

\bigskip

Note that by construction,
    $\;(A_{\varphi})_{\redscriptsize}\,
	        \simeq\, A_{\varphi_{\redtiny}}\,
			=\, C^{\infty}((X_{\varphi})_{\redscriptsize})\;$
  canonically.
   
\bigskip

\begin{scorollary} {\bf [masslessness of $\nabla$ viewed from open string]}$\;$
 The connection $\nabla$ on $E$ is massless viewed from open string in $Y$ via $\varphi$.
\end{scorollary}

\smallskip

\noindent
{\it Reason.}
 The answer to whether a field on the D-brane world-volume is massless or massive from the viewpoint of open strings
   is determined by how that field is created as part of the spectrum of excitations of open strings.
 As the string tension is fixed, the only way an open string can create a massless field on the D-brane world-volume
  through excitations is
   when that open string on one hand has its both end-points on that D-brane and on the other hand has its length nearly zero.
 
 In our setting, an open string is moving in $Y$ and it interacts with the D-brane world-volume $X^{\!A\!z}$ via
   the image $\varphi(X^{\!A\!z})$ of $X^{\!A\!z}$ in $Y$.
 Recall the map $f_{\varphi}:X_{\varphi}\rightarrow Y$ associated to $\varphi$.
 For any $x\in X$, the restriction
  $f_{\varphi}|_{\pi_{\varphi}^{-1}(x)}: \pi_{\varphi}^{-1}(x)\rightarrow Y$
  is an embedding.
 That
   there exists a decomposition
    $({\cal E}_x,\nabla)= \oplus_{q\in\pi_{\varphi}^{-1}(x)}(e_q{\cal E},\, ^q\nabla)$
  says that there are no components of $\nabla$ that arise from excitation of an open string
  with one end on some $f_{\varphi}(q)$ and the other end on some $f_{\varphi}(q^{\prime})$,
  $q^{\prime}\ne q$.
 It follows that $\nabla$ is massless from the viewpoint of open strings.
     
\hspace{39.5em}$\square$

\bigskip

\noindent
Cf.\ {\sc Figure}~2-2.
%

\begin{figure}[htbp]
 \bigskip
  \centering
  \includegraphics[width=0.80\textwidth]{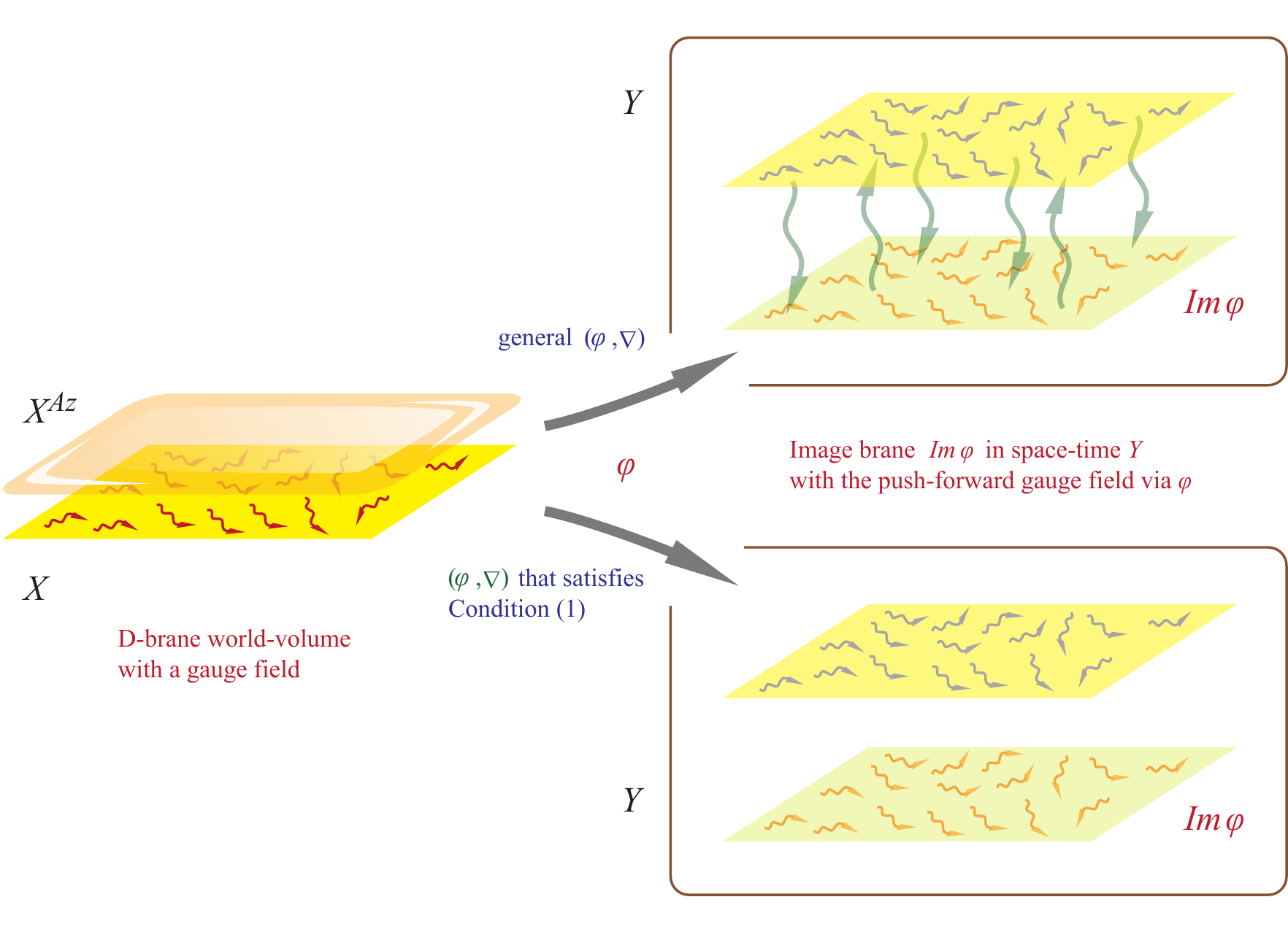}
 
  \bigskip
  \bigskip
 \centerline{\parbox{13cm}{\small\baselineskip 12pt
  {\sc Figure}~2-2.
  As the tension of open strings are constant, the mass of an open string is proportional to its length.
  Nearly massless open strings create massless fields on a D-brane world-volume
     where the two end-points of open strings lie in,
    while massive open strings create massive ones.
  Thus, excitations of an open string of length nearly zero create massless fields on the D-brane world-volume
   while excitations of an open string of length bounded below by a positive number create massive ones.
  The decomposition
   $$
    ({\cal E}_x,\,\nabla)\;
	   =\; \oplus_{q\in\pi_{\varphi}^{-1}}\,(e_q\,{\cal E}_x,\,^q\nabla)\,,
   $$
   a consequence of Condition (1) on $(\varphi,\nabla)$,
  says that the gauge field $\nabla$ on any small neighborhood $V$ of $x\in X$
   localizes at each connected branch of $\varphi(V^{\!A\!z})$ from the viewpoint of open strings in $Y$.
 This says that
    for $(\varphi,\nabla)$ satisfying Condition (1),  $\nabla$ is massless from the open-string aspect.
 In the illustration,
  the noncommutative space $X^{\!A\!z}$ is expressed as a noncommutative cloud shadowing over its underlying topology $X$,
  the connection $\nabla$ on $E$ over $X$  is indicated by a gauge field on $X$.
 Both the gauge field on $X$   and how open strings ``see" it in $Y$
  are indicated by squiggling arrows $\rightsquigarrow$.
 The situation for a general $(\varphi,\nabla)$  (cf.\ top) and a $(\varphi,\nabla)$ satisfying Condition (1) (cf.\ bottom)
  are compared.
 From the open-string aspect, in the former situation $\nabla$ can have both massless components
  (which are local fields from the open-string and target-space viewpoint) and massive components
  (which become nonlocal fields from the open-string and target-space viewpoint),
  while in the latter situation $\nabla$ has only massless components.
  }}
\end{figure}

\bigskip

\section{Additional Condition (2)}

Assuming that Condition (1) holds,
we now examine the additional Condition (2) on pairs $(\varphi,\nabla)$:
 $$
    F_{\nabla}\;
	    \subset\;  C^{\infty}(\Omega_X^2)\otimes_{C^{\infty}(X)} \Comm(A_{\varphi})\,.
 $$		
Before proceeding, we state two lemmas that are elementary but play the key role:

\bigskip

\begin{slemma} {\bf [product with nilpotent element]}$\;$
 Let $R$ be a (possibly noncommutative) ring and $n\in R$ be nilpotent.
 Let $r_1,\,\cdots\,,r_l$ be arbitrary elements in $R$ that commute with $n$.
 Then any product of the form
  $$
    r_{\sigma(1)}\,\cdots\,r_{\sigma(k)}\,n\,r_{\sigma(k+1)}\,\cdots\,r_{\sigma(l)}
  $$
  is nilpotent.
 Here, $\sigma\in\Sym_{(l)}$ is a permutation.
\end{slemma}

\bigskip

Note that the statement of false without the condition that $r_1,\,\cdots\,,\, r_l$ commute with $n$.
(Recall [L-Y: Sec.\ 3.1.4; in particular, proof of Lemma 3.1.4.3] for omitted explanation in the next lemma.).

\bigskip

\begin{slemma} {\bf [square root of reduced component]}$\;$
 Let
   $m\in \GL(r,{\Bbb C})\in M_{r\times r}({\Bbb C})$ be an invertible $r\times r$ matrix  and
   $m\;=\; m_0 \,+\,n$ be the decomposition of $m$ into the diagonalizable component $m_0$
    of the nilpotent component $n$.
 In terms of the finite-dimensional algebra generated by $m$,
   this decomposition coincide with the decomposition of $m$ into the reduced component, i.e.\ $m_0$,
       plus the nilpotent component, i.e.\ $n$; cf.\ Lemma~2.6 and Definition~2.7.
  Then,
     $$
	    \sqrt{m}\;=\; \sqrt{m_0} \, +\, \mbox{\it nilpotent}\,.
	 $$
   Here, $\sqrt{\tinybullet}$ is the principal square root map.
\end{slemma}

\bigskip

\begin{flushleft}
{\bf Simplification of $S_{\DBI}(\varphi,\nabla)+S_{\CSWZ}(\varphi,\nabla)$}
\end{flushleft}
Let
 $(\varphi,\nabla)$ satisfy Condition (1).
Take $X=U$ and let
 $$
   \varphi^{\sharp}\;=\; \varphi^{\sharp}_{\redscriptsize}\,+\, \xi_{\varphi}
 $$
 be the canonical decomposition of
 $\varphi^{\sharp}: C^{\infty}(Y) \rightarrow C^{\infty}(\End_{\Bbb C}(E))$
 from Lemma~2.6.
	
\bigskip

\begin{sproposition}
{\bf [additional Condition (2) $\Rightarrow$ nilpotent component of $\varphi^{\sharp}$  decoupled]}$\;$
 If, in addition to Condition (1), Condition (2) is also satisfied,
 then
  $$
    S_{\DBI}(\varphi,\nabla)\; =\; S_{\DBI}(\varphi_{\redscriptsize},\nabla)
	     \hspace{2em}\mbox{and}\hspace{2em}
    S_{\CSWZ}(\varphi,\nabla) \; =\;  S_{\CSWZ}(\varphi_{\redscriptsize},\nabla) \,.	
  $$
 Here, it is assumed that
   the anomaly factor in the integrand of the Chern-Simons/Wess-Zumino term $S_{\CSWZ}$ is $1$.
\end{sproposition}

\smallskip

\begin{proof}
 In a local chart $U^{\prime}$ of $X$,
  the Dirac-Born-Infeld action $S_{\DBI}(\varphi,\nabla)$,
   with the target-space(-time)  $Y$ equipped with metric tensor $g$, $B$-field $B$, and a dilaton field $\Phi$,
   is given by
  
 {\footnotesize
  \begin{eqnarray*}
   \lefteqn{
    S_{\DBI}^{(\Phi,g,B)}(\varphi|_{U^{\prime}},\nabla|_{U^{\prime}})  }\\
    &&	=\;
       \mp\,T_{m-1}\,\int_{U^{\prime}}
   	                       \Tr
						    e^{-\varphi^{\sharp}(\Phi)}
						    \sqrt{\mp\,
						      \SymDet_U \left( \rule{0ex}{1em} \right.
		                         \sum_{i,j}
								    \varphi^{\sharp}(g_{ij}+ B_{ij})\,
								      D_{\mu}\varphi^{\sharp}(y^i)\,D_{\nu}\varphi^{\sharp}(y^j)\,
		                              +\,  2\pi\alpha^{\prime}\, F_{\mu\nu}
                                       \left. \rule{0ex}{1em}	                                    
 										\right)_{\mu\nu} }\;   d^{\,m}x\,.	 	
  \end{eqnarray*}}

 \noindent
 Recall that
    $e^{-\varphi^{\sharp}(\Phi)}$ and
	all $\varphi^{\sharp}(g_{ij}+ B_{ij})$,
	     $D_{\mu}\varphi^{\sharp}(y^i)$, $D_{\nu}\varphi^{\sharp}(y^j)$
   are in $A_{\varphi}|_{U}$.
 One then has
   \begin{eqnarray*}
    \lefteqn{
       \sum_{i,j}\varphi^{\sharp}(g_{ij}+ B_{ij})\,
							D_{\mu}\varphi^{\sharp}(y^i)\,D_{\nu}\varphi^{\sharp}(y^j)}\\
       &&  =\;
	             \sum_{i,j}
					\varphi_{\redscriptsize}^{\sharp}(g_{ij}+ B_{ij})\,
				      D_{\mu}\varphi_{\redscriptsize}^{\sharp}(y^i)\,
					  D_{\nu} \varphi_{\redscriptsize}^{\sharp}(y^j)\;
                      +\, (\mbox{\it nilpotent terms})					
   \end{eqnarray*}							

 When Condition (2) is satisfied,
  $F_{\mu\nu}$'s commute with elements of $A_{\varphi}$
  and it follows from Lemma~3.1 and Lemma~3.2 that
 %
 {\footnotesize
 \begin{eqnarray*}
  \lefteqn{
    \sqrt{\mp\,
						      \SymDet_U \left( \rule{0ex}{1em} \right.
		                         \sum_{i,j}
								    \varphi^{\sharp}(g_{ij}+ B_{ij})\,
								      D_{\mu}\varphi^{\sharp}(y^i)\,D_{\nu}\varphi^{\sharp}(y^j)\,
		                              +\,  2\pi\alpha^{\prime}\, F_{\mu\nu}
                                       \left. \rule{0ex}{1em}	                                    
 										\right)_{\mu\nu} }   }\\
   && =\; \sqrt{\mp\,
						      \SymDet_U \left( \rule{0ex}{1em} \right.
		                         \sum_{i,j}
								    \varphi_{\redtiny}^{\sharp}(g_{ij}+ B_{ij})\,
								      D_{\mu}\varphi_{\redtiny}^{\sharp}(y^i)\,
									  D_{\nu}\varphi_{\redtiny}^{\sharp}(y^j)\,
		                              +\,  2\pi\alpha^{\prime}\, F_{\mu\nu}
                                       \left. \rule{0ex}{1em}	                                    
 										\right)_{\mu\nu}\;
                                    +\; \mbox{\it nilpotent}\;\;										}	\\
   && =\; \sqrt{\mp\,
						      \SymDet_U \left( \rule{0ex}{1em} \right.
		                         \sum_{i,j}
								    \varphi_{\redtiny}^{\sharp}(g_{ij}+ B_{ij})\,
								      D_{\mu}\varphi_{\redtiny}^{\sharp}(y^i)\,
									  D_{\nu}\varphi_{\redtiny}^{\sharp}(y^j)\,
		                              +\,  2\pi\alpha^{\prime}\, F_{\mu\nu}
                                       \left. \rule{0ex}{1em}	                                    
 										\right)_{\mu\nu}\;}\;\;
                                        +\;\; \mbox{\it nilpotent}^{\prime}																		
 \end{eqnarray*}}
 
 Together with
   $$
      e^{-\varphi^{\sharp}(\Phi)}\;
	    =\; e^{-\varphi_{\redtiny}^{\sharp}(\Phi)}\,+\, \mbox{\it nilpotent}
   $$
   and
   $$
     \Tr(\,\mbox{\it nilpotent matrix}\,)\;=\; 0\,,
   $$
   one has
  
 {\footnotesize
  \begin{eqnarray*}
   \lefteqn{
    S_{\DBI}^{(\Phi,g,B)}(\varphi|_{U^{\prime}},\nabla|_{U^{\prime}})  }\\
    &&	=\;
       \mp\,T_{m-1}\,\int_{U^{\prime}}
   	                       \Tr
						    e^{-\varphi_{\redtiny}^{\sharp}(\Phi)}
						    \sqrt{\mp\,
						      \SymDet_U \left( \rule{0ex}{1em} \right.
		                         \sum_{i,j}
								    \varphi_{\redtiny}^{\sharp}(g_{ij}+ B_{ij})\,
								      D_{\mu}\varphi_{\redtiny}^{\sharp}(y^i)\,
									  D_{\nu} \varphi_{\redtiny}^{\sharp}(y^j)\,
		                              +\,  2\pi\alpha^{\prime}\, F_{\mu\nu}
                                       \left. \rule{0ex}{1em}	                                    
 										\right)_{\mu\nu} }\;   d^{\,m}x	 	 \\
    &&  =\;
	       S_{\DBI}^{(\Phi,g,B)}(\varphi_{\redtiny}|_{U^{\prime}},\nabla|_{U^{\prime}})\,.
  \end{eqnarray*}}
 This proves the identity for $S_{\DBI}$.
 
 Similarly, for the identity for the Chern-Simons/Wess-Zumino term $S_{\CSWZ}$.

 This proves the proposition.

\end{proof}
 
\bigskip

\noindent
In other words,
  the true dynamical fields in the theory are now $(\varphi_{\redscriptsize},\nabla)$
   and  the nilpotent component $\xi_{\varphi}$ of $\varphi^{\sharp}$
   serves pretty much like an auxiliary field in the problem:
 Together with $\varphi_{\redscriptsize}$,
   it imposes algebraic constraints on $\nabla$ but is itself non-dynamical and decouples locally from the theory.
 Cf.\ Remark~3.5.

\bigskip

\begin{sremark} $[$Condition (1) alone$]\;$ {\rm
 If $(\varphi,\nabla)$ satisfies only Condition (1),
   then both $\varphi^{\sharp}_{\redscriptsize}$ and $\xi_{\varphi}$ are dynamical.
 The three $\varphi^{\sharp}_{\redscriptsize}$, $\xi_{\varphi}$, and $\nabla$ are coupled with
  and influence each other.
 The dynamics of D-branes is conceivably much more complicated, albeit much more interesting.
}\end{sremark}
   
\medskip

\begin{sremark} $[$global effect$]\;$ {\rm
 When $(\varphi,\nabla)$ satisfies both Condition (1) and Condition (2),
    though $\xi_{\varphi}$ is non-dynamical,
  it may still have some global effect to D-brane dynamics to be understood.
 Furthermore, when $U\subsetneqq X$,
  the singular locus of $X_{\varphi}$ over $X$ can have effect to the dynamics of
    $(\varphi_{\redscriptsize},\nabla)$ as well since $(f_{\redscriptsize})$ must extend to a continuous map
	$(X_{\varphi})_{\redscriptsize}\rightarrow Y$
	on the whole presumably-singular-over-$(X-U)\,$ $X_{\varphi}$.
}\end{sremark}

\bigskip

\begin{flushleft}
{\bf A refinement of admissibility}
\end{flushleft}
The study in Sec.\ 2 of this note  suggests the following refined definition for future use.

\bigskip

\begin{sdefinition} {\bf [weakly vs.\ strongly admissible pair $(\varphi,\nabla)$]}$\;$ {\rm
 Let $\varphi:(X^{\!A\!z},E)\rightarrow Y$ be a differentiable map  and
  $\nabla$ be a connection on $E$.
 The pair $(\varphi,\nabla)$ is called {\it weakly admissible} if Condition (1) is satisfied
  over an open-dense subset of $X$:
    \begin{itemize}
     \item[(1)]\hspace{3em}
       $DA_{\varphi}\;\subset\; C^{\infty}(\Omega_X)\otimes_{C^{\infty}(X)}A_{\varphi}$.
	\end{itemize}
 And is called {\it admissible}	 or {\it strongly admissible} if, in addition, Condition (2) is also satisfied:
	\begin{itemize}
	 \item[(2)]\hspace{3em}
	   $F_{\nabla}\;
	      \subset\;  C^{\infty}(\Omega_X^2)\otimes_{C^{\infty}(X)} \Comm(A_{\varphi})\,.$
   \end{itemize}
}\end{sdefinition}

\bigskip

\begin{flushleft}
{\bf The anomaly factor in the Chern-Simons/Wess-Zumino term $S_{\CSWZ}(\varphi,\nabla)$}
\end{flushleft}
It follows from Lemma~2.8 that
 when
   $$
     \varphi\;:\; (X^{\!A\!z},E;\nabla)\; \longrightarrow \; Y
   $$
    satisfies Condition (1) (i.e.\ weakly admissible),
 $(E,\nabla)$ induces
   a bundle $(_{\varphi_{\redtiny}}\!E,\,  _{\varphi_{\redtiny}}\!\!\nabla) $ with a connection
   on $(X_{\varphi})_{\redscriptsize}$ over an open dense subset $U$ of $X$.
And, over $U$, there is map
  $$
     \varphi_{\redscriptsize}\;:\; (X^{\!A\!z},E;\nabla)\; \longrightarrow \; Y
  $$
 whose defining ring-homomorphism
  $\varphi_{\redscriptsize}^{\,\sharp}:
                     C^{\infty}(Y)\rightarrow  C^{\infty}(\End_{\Bbb C}(E))$
 differs from $\varphi^{\sharp}$ by a derivation
   $\xi_{\varphi}:  C^{\infty}(Y)\rightarrow  C^{\infty}(\End_{\Bbb C}(E))$
   that takes values in nilpotent elements of $A_{\varphi}$.

Let
 $$
    \breve{X}\; :=\;  (X_{\varphi})_{\redscriptsize}|_U
	 \hspace{2em}\mbox{and}\hspace{2em}
  (\breve{E}, \breve{\nabla})\;
	   :=\; (_{\varphi_{\redtiny}}\!E,\,  _{\varphi_{\redtiny}}\!\!\nabla)\,.
 $$
Then
  through the built-in inclusions
  $A_{\varphi_{\redtiny}}|_U
      \subset C^{\infty}(\End_{\Bbb C}(\breve{E}))\subset C^{\infty}(\End_{\Bbb C}(E))$,
 $\varphi_{\redscriptsize}^{\,\sharp}$ induces canonically a ring-homomorphism
  $$
    \breve{\varphi}^{\sharp}\;:\;  C^{\infty}(Y)\;
	   \longrightarrow\; C^{\infty}(\End_{\Bbb C}(\breve{E}))
  $$
 over ${\Bbb  R}\hookrightarrow {\Bbb C}$,
  which defines a map
   $$
    \breve{\varphi}\;:\;  (\breve{X}^{\!A\!z}, \breve{E};\breve{\nabla})\;\longrightarrow\; Y\,.
   $$
By construction,
 $$
   A_{\breve{\varphi}}\; =\; C^{\infty}(\breve{X})
 $$
  through the built-in inclusion
  $C^{\infty}(\breve{X})\subset C^{\infty}(\End_{\Bbb C}(\breve{E}))$,
which means that
 $\breve{\varphi}$ really comes from a map
  $$
    \breve{f}\;:\; \breve{X}\;\longrightarrow\;  Y\,.
  $$
And one has the following commutative diagram over $U$
 $$
   \xymatrix{
    X^{\!A\!z}   \ar@<+.4ex>[rrrrr]^-{\varphi_{\redtiny}} \ar@{->>}[rd]
	                          \ar@{->>}[d]_-{\sigma_{\varphi_{\redtiny}}} &&&&& Y \\
	X_{\varphi_{\redtiny}}  \ar@/^2ex/[rrrrru]^(.35){f_{\varphi_{\redtiny}}}
	                                |(.22){\mbox{\rule{0ex}{1ex}$\hspace{3em}$}}
								  \ar@{=}[rd]	
								  \ar@{->>}[d]_-{\pi_{\varphi_{\redtiny}}}
	    &   \breve{X}^{\!A\!z}\ar[rrrru]^-{\breve{\varphi}}
		                             \ar@{->>}[d]^-{\breve{\sigma}_{\breve{\varphi}}}  \\
    X
	    & \rule{0ex}{1em} \hspace{6ex}\breve{X}_{\breve{\varphi}}\hspace{6ex}
		            \ar@/^1ex/[rrrruu]_-{\breve{f}_{\breve{\varphi}} }
                     \ar@{->>}[dd]|{\hspace{1ex}\breve{\pi}_{\breve{\varphi}}\;=\; I\!d_{\breve{X}}}\\	\\
		& \breve{X} \ar@{->>}[uul]^-{\pi_{\breve{\varphi}}}  \ar[rrrruuuu]_-{\breve{f}}
   }
 $$	
 that relates all the built-in objects and maps underlying $\varphi_{\redscriptsize}$ and $\breve{\varphi}$   and
        extends the previous diagram in Sec.\ 2 relating $\varphi$ and $\varphi_{\redscriptsize}$ over $U$.

If, in addition,  $(\varphi, \nabla)$ satisfies Condition (2) (i.e.\ (strongly) admissible),
then it follows from Corollary~3.3 and the above diagram that
 $$
   S_{\DBI}(\varphi,\nabla)\; =\; S_{\DBI}(\varphi_{\redscriptsize},\nabla)\;
	=\; S_{\DBI}(\breve{f},\breve{\nabla})
 $$
  and
 $$
   S_{\CSWZ}(\varphi,\nabla) \; =\;  S_{\CSWZ}(\varphi_{\redscriptsize},\nabla)\;
    =\; S_{\CSWZ}(\breve{f},\breve{\nabla})	\,.	
 $$
Again, it is assumed that
   the anomaly factor in the integrand of the Chern-Simons/Wess-Zumino term $S_{\CSWZ}$ is $1$.

However,
  in the current situation and under the requirement that $\varphi$ is either Lorentzian or Riemannian
  (cf.\ [L-Y: Definition 3.1.2.1, Definition 3.1.2.2, Definition 3.2.1] (D(13.1))),
  one has
 \begin{itemize}
  \item[\LARGEdot]{\it
    The map $\breve{f}:\breve{X}\rightarrow Y$ is indeed an immersion.}
 \end{itemize}	
{\it Assume further that $U=X$ is closed.}
 Then it is known that in this case
  the anomaly factor in the integrand of the Chern-Simons/Wess-Zumino term $S_{\CSWZ}$ is given by
 (e.g.\ [AG-G], [AG-W], [C-Y], [G-M-H], [K-Y], [S-S])
  $$
    \mbox{\it AnomalyFactor}\,(\breve{f})\;
	 =\; \sqrt{\hat{A}(T_{\ast}\breve{X})/\hat{A}(N_{\breve{f}})\:}
	        \:  \in\; \Omega^{\ast}(\breve{X})\,.
  $$
 Here,
  $\hat{A}(\,\cdot\,)$ is the $\hat{A}$-class of the bundle in question,
  $N_{\breve{f}}$ is the normal bundle of $\breve{X}$ in $Y$ along $\breve{f}$,
  $\Omega^{\ast}(\breve{X})$
    is the ${\Bbb Z}$-graded-commutative $C^{\infty}(\breve{X})$-module
    of differential forms on $\breve{X}$,  and
  the class
    $\sqrt{\hat{A}(T_{\ast}\breve{X})/\hat{A}(N_{\breve{f}})\:}$
	is represented by a differential form on $\breve{X}$.
With the built-in inclusions
 $C^{\infty}(\breve{X})\subset A_{\varphi}\subset  C^{\infty}(\End_{\Bbb C}(E))$,
 one may identify $\mbox{\it AnomalyFactor}(\breve{f})$ canonically
   with an $\End_{\Bbb C}(E)$-valued differential form on $X$, in notation,
   $\pi_{\breve{\varphi},\ast}\,(\mbox{\it AnomalyFactor}\,(\breve{f}))$.
Since
 $\;S_{\CSWZ}(\varphi,\nabla) = S_{\CSWZ}(\varphi_{\redscriptsize},\nabla)
    =S_{\CSWZ}(\breve{f},\breve{\nabla})\;$
 when the anomaly factor in the integrand of the Chern-Simons/Wess-Zumino term $S_{\CSWZ}$ is set to $1$,
it is suggestive that, possibly up to a nilpotent-$\End_{\Bbb C}(E)$-valued differential form
       (which will decouple after taking $\int_X(\,\cdots\,)$),
 one has
 $$
   \mbox{\it AnomalyFactor}\,(\varphi)\;
     =\;  \mbox{\it AnomalyFactor}\,(\varphi_{\redscriptsize})\;
	 =\;  \pi_{\breve{\varphi},\ast}\,(\mbox{\it AnomalyFactor}\,(\breve{f}))
 $$
 and can define the Chern-Simons/Wess-Zumino term for $(\varphi,\nabla)$ as
 {\small
 $$
  S_{\CSWZ}^{(C,B)}(\varphi,\nabla)\;
   =\;
     T_{m-1}\,\int_X  \Real\left( \Tr
    	 \left(\varphi^{\diamond}C
		            \odotwedge e^{2\pi\alpha^{\prime}F_{\nabla}+\varphi^{\diamond}B}
	                \odotwedge
					\pi_{\breve{\varphi},\ast}\sqrt{\hat{A}(T_{\ast}\breve{X})/\hat{A}(N_{\breve{f}})\,}
					\right) \right)_{(m)}\,.
 $$}
Here,
 $m$ is the dimension of $X$  and
 $(B,C)$ is a background $(\mbox{$B$-field, Ramond-Ramond field})$ on the target $Y$, and
 $\odotwedge$ is the symmetrized wedge-product for $\End_{\Bbb C}(E)$-valued differential forms on $X$.
See  [L-Y: Sec.\ 6.1] (D(13.1)) for a more complete explanation.

With this modification of the anomaly factor, it remains to hold that
 for $(\varphi,\nabla)$ (strongly) admissible,
 $$
  S_{\CSWZ}^{(C,B)}(\varphi,\nabla)\;
  =\; S_{\CSWZ}^{(C,B)}(\varphi_{\redscriptsize},\nabla)\,.
 $$
 
When $U\subsetneqq X$, $U$ is noncompact and has boundary in $X$.
How to define $\mbox{\it AnomalyFactor}(\varphi)$
 or $\mbox{\it AnomalyFactor}(\varphi_{\redscriptsize})$
 in such a situation remains to be understood.

\newpage
\baselineskip 13pt
{\footnotesize

\vspace{1em}

\noindent
chienhao.liu@gmail.com, chienliu@math.harvard.edu; \\
yau@math.harvard.edu

}

\end{document}